\def\year{2022}\relax
\documentclass[hidelinks,letterpaper]{article} 
\usepackage{aaai20} 
\usepackage{times}  
\usepackage{helvet} 
\usepackage{courier}  
\usepackage[hyphens]{url}  
\usepackage{graphicx} 
\usepackage{booktabs}
\usepackage{mathtools}
\usepackage{multirow}
\usepackage{multicol}
\usepackage{color, colortbl}
\usepackage[dvipsnames]{xcolor}
\usepackage{amssymb}
\usepackage{subfig}
\usepackage{algorithmic}
\usepackage{todonotes}
\usepackage{bigdelim}
\usepackage{comment}
\usepackage{tikz}
\usepackage{hyperref}
\usepackage{amsmath}

\DeclareMathOperator*{\argmax}{arg\,max}

\usepackage{pgfplots, pgfplotstable}
\usepgfplotslibrary{colormaps}%
\pgfplotsset{compat=1.15}
\pgfplotsset{
    colormap={slategraywhite}{
        rgb255=(112,128,144)
        rgb255=(255,159,101)
    }}
\usepackage{graphicx}  
\definecolor{Gray}{gray}{0.9}
\definecolor{Blue}{gray}{0.7}
\def\RemoveSpaces#1{\zap@space#1 \@empty}
\newcommand{\citet}[1]{\citeauthor{#1} \shortcite{#1}}

\title{
{\fontfamily{cmss}\selectfont
ISEEQ:
} Information Seeking Question Generation using Dynamic Meta-Information Retrieval and Knowledge Graphs}
\author{Manas Gaur\thanks{Research work was done while author was interning at Samsung Research America.}\textsuperscript{$\dagger$},
 Kalpa Gunaratna\textsuperscript{$\ddagger$},
 Vijay Srinivasan\textsuperscript{$\ddagger$},
 Hongxia Jin\textsuperscript{$\ddagger$}
\\
\textsuperscript{$\dagger$}{AI Institute, University of South Carolina, SC, USA }\\ {mgaur@email.sc.edu}\\
\textsuperscript{$\ddagger$}{Samsung Research America, Mountain View, CA, USA}\\
\{k.gunaratna, v.srinivasan, hongxia.jin\}@samsung.com}

\pgfplotstableread[col sep=comma,]{data.csv}\datatable

\pgfplotsset{colormap/violet}\pgfplotstableread[col sep=comma,]{data2.csv}\datatables

\begin{document}
\maketitle
\begin{abstract}

Conversational Information Seeking (CIS) is a relatively new research area within conversational AI that attempts to seek information from end-users in order to understand and satisfy users' needs. If realized, such a system has far-reaching benefits in the real world; for example, a CIS system can assist clinicians in pre-screening or triaging patients in healthcare. A key open sub-problem in CIS that remains unaddressed in the literature is generating Information Seeking Questions (ISQs) based on a short initial query from the end-user. To address this open problem, we propose \textbf{I}nformation \textbf{SEE}king \textbf{Q}uestion generator {\fontfamily{cmss}\selectfont (ISEEQ)}, a novel approach for generating ISQs from just a short user query, given a large text corpus relevant to the user query.  Firstly, {\fontfamily{cmss}\selectfont ISEEQ} uses a knowledge graph to enrich the user query.
Secondly, {\fontfamily{cmss}\selectfont ISEEQ} uses the knowledge-enriched query to retrieve relevant context passages to ask coherent ISQs adhering to a conceptual flow. Thirdly, {\fontfamily{cmss}\selectfont ISEEQ} introduces a new deep generative-adversarial reinforcement learning-based approach for generating ISQs. We show that {\fontfamily{cmss}\selectfont ISEEQ} can generate high-quality ISQs to promote the development of CIS agents. {\fontfamily{cmss}\selectfont ISEEQ} significantly outperforms comparable baselines on five ISQ evaluation metrics across four datasets having user queries from diverse domains. Further, we argue that {\fontfamily{cmss}\selectfont ISEEQ} is transferable across domains for generating ISQs, as it shows the acceptable performance when trained and tested on different pairs of domains. The qualitative human evaluation confirms {\fontfamily{cmss}\selectfont ISEEQ}-generated ISQs are comparable in quality to human-generated questions and outperform the best comparable baseline.

\end{abstract}

\section{Introduction}
\label{sec:intro}
Information Seeking (IS) is a complex and structured process in human learning that demands lengthy discourse between seekers and providers to meet the information needs of the seekers. The provider can ask the seeker information-seeking questions to understand the seeker's needs better and respond appropriately. For instance, clinicians use their experience or medical knowledge to ask patients information-seeking questions (ISQs), who describe their health condition (a short initial IS query). Conversational Information Seeking (CIS) is an emerging research area within conversational AI that aims to emulate the provider by automatically asking ISQs, keeping track of seeker responses, and ultimately responding to the seeker's needs based on responses to ISQs. CIS has broadened the research scope of various virtual assistants (e.g., Alexa, Bixby)~\cite{zamani2020macaw,radlinski2017theoretical}. Existing work in the area of CIS has primarily focused on aspects such as retrieving relevant passages to respond to seeker queries and generating answers \cite{vakulenko2021large,kumar2020making}. 

\begin{figure}[t]
  \begin{center}
    \includegraphics[width=85mm, scale=0.30, trim=0.cm 1.6cm 0.0cm 0.8cm]{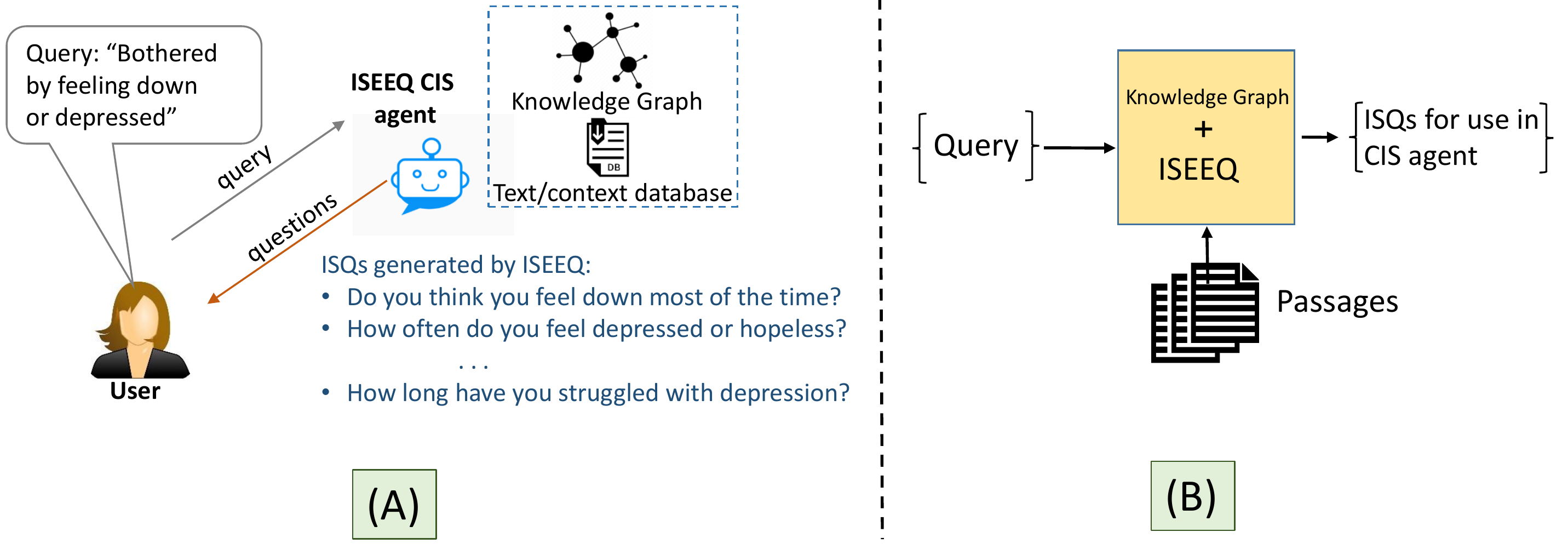}
  \end{center}
\caption{\footnotesize (A) An example of curiosity-driven ISQs generated by {\fontfamily{cmss}\selectfont ISEEQ}. (B) overview of {\fontfamily{cmss}\selectfont ISEEQ}}
\label{fig:intro_figure}
\end{figure}

To the best of our knowledge, the problem of generating ISQs given an initial IS query from the user has not been addressed in the literature so far. Figure \ref{fig:intro_figure}A shows example ISQs generated for the user IS query ``Bothered by feeling down or depressed''. For example, user responses to ISQs such as ``How often do you feel depressed or hopeless?'' and ``How long have you struggled with depression?'' can be used either by the CIS or the healthcare provider to generate an appropriate response to the user's needs. ISQs differ from other question types (e.g., Clarifying questions, Follow-up questions \cite{rao2018learning,zamani2020generating,pothirattanachaikul2020analyzing}) by having a structure, covering objective details, and expanding on the breadth of the topic. For such a structure between ISQs, there are semantic relations and logical coherence (together termed as \textit{conceptual flow}). From \cite{aliannejadi2019asking}, clarifying questions are simple questions of facts, good to clarify the dilemma, and \textit{confined to the entities in the query}. In contrast, ISQs go a step further with expanding the query context by \textit{exploring relationships between entities in the query and linked entities in a knowledge graph}. Thus retrieving a diverse set of passages that would provide a proper solution to a user query.

Firstly, a key challenge in generating ISQs is that the initial IS query is short and has limited context. Without explicit integration of external knowledge for enriching the IS query, CIS cannot achieve the curiosity-driven generation of ISQs. Secondly, training an ISQ generation system requires annotated datasets containing IS queries, ISQs, and many passages. Creating such datasets requires skilled and trained crowdsource workers~\cite{dalton2020cast}. Moreover, the process is (i) tedious for the crowd worker in terms of the number of passages needed for question creation and can result in (ii) insufficient question coverage when the answer to a query lies across multiple passages, requiring workers to perform extensive search \cite{wood2020dialogue,wood2018detecting}.

To address these challenges and the open problem of ISQ generation,  we present \textbf{I}nformation \textbf{SEE}king \textbf{Q}uestion generator {\fontfamily{cmss}\selectfont (ISEEQ)} to enable a curiosity-driven CIS system. Essentially, the design of {\fontfamily{cmss}\selectfont ISEEQ} relies on exploring three research questions: \textbf{(RQ1)} \textit{Knowledge-infusion}: Can expert-curated knowledge sources like knowledge graphs/bases related to the user query help in context retrieval and question generation? \textbf{(RQ2)} \textit{conceptual flow}: Can {\fontfamily{cmss}\selectfont ISEEQ} generate ISQs having semantic relations and logical coherence? \textbf{(RQ3)} Can {\fontfamily{cmss}\selectfont ISEEQ} generate ISQs in a cross domain setting and generate ISQs for new domains without requiring crowdsourced data collection?. We believe addressing the three \textbf{RQs} uniquely positions this research as the first to develop a successful solution to ISQ generation for CIS. Figure \ref{fig:intro_figure}B shows the overall inputs and outputs of ISEEQ. ISEEQ generates ISQs based on a short IS query from the seeker, by making use of a large text corpus of passages relevant to the IS query and also relevant knowledge graphs.

Our key contributions of this work are as follows:

\begin{enumerate}
    \item \textbf{Problem definition and approach:} To the best of our knowledge, we are the first to formulate the problem of automatic generation of ISQs for CIS. To solve this, we introduce a novel approach called {\fontfamily{cmss}\selectfont ISEEQ} that can \textit{automatically} generate curiosity-driven and conceptual flow-based ISQs from a short user query. 
    
    \item \textbf{Dynamic knowledge-aware passage retrieval:} We infuse IS queries with semantic information from knowledge graphs to improve unsupervised passage retrieval. Passages serve as meta-information for generating ISQs. 
   
    \item \textbf{Reinforcement learning for ISQs:} To improve compositional diversity and legibility in QG, we allow {\fontfamily{cmss}\selectfont ISEEQ} self-guide the generations through reinforcement learning in generative-adversarial setting that results in {\fontfamily{cmss}\selectfont ISEEQ-RL}. We introduce entailment constraints borrowed from natural language inference (NLI) guidelines to expand {\fontfamily{cmss}\selectfont ISEEQ-RL} to {\fontfamily{cmss}\selectfont ISEEQ-ERL} to have smooth topical coherent transitions in the questions, achieving conceptual flow.
    \item \textbf{Evaluation metrics:} We introduce two evaluation metrics: ``semantic relations'' and ``logical coherence'' to measure conceptual flow in the generated questions.
    
\end{enumerate}

We evaluated {\fontfamily{cmss}\selectfont ISEEQ} (both {\fontfamily{cmss}\selectfont ISEEQ-RL} \& {\fontfamily{cmss}\selectfont ISEEQ-ERL}) using four conversational discourse datasets with five natural language generation metrics. In quantitative evaluation, {\fontfamily{cmss}\selectfont ISEEQ} shows superiority over state-of-the-art approaches considered for CIS. We show that {\fontfamily{cmss}\selectfont ISEEQ} is transferable across domains for generating ISQs, as it shows acceptable performance when trained and tested on different pairs of domains; this addresses the key challenge of reducing human effort in training ISQ generation models for new domains. Moreover, 12 human evaluations of 30 IS queries show that {\fontfamily{cmss}\selectfont ISEEQ} generated ISQs are comparable to ground truth human generated questions and they outperformed a competitive baseline generated ones.

\section{Related Work}
\label{sec:related}
CIS understands that conversations possess a well-defined structure that addresses the information needs of the user initiating the conversation \cite{li2021conversations}. The datasets to train models in CIS are designed to facilitate a mixed-initiative dialogue, where the agent can also ask clarifying or follow-up questions to gather information concerning the IS query \cite{zhang2019addressing}. Alongside, datasets like ShARC help CIS models to improve response generation \cite{saeidi2018interpretation}. However, in this study, we restrict ourselves to solving an open sub-problem in CIS that generates ISQs. 
At present, passage retrieval and ranking, turn-by-turn question generation, search, and answer retrieval have been independently explored within CIS \cite{vakulenko2021large}. Dense Passage Retrieval (DPR) retrieves and ranks passages using maximum inner product search (MIPS), using a dense representation based task-specific similarity function. QuAC, HOTPOTQA, WikipassageQA and Topical Chat were introduced to train \textit{IS dialogue} for non-factoid question answering from the retrieved passages \cite{choi2018quac,yang2018hotpotqa,Gopalakrishnan2019,cohen2018wikipassageqa}. Together with DPR, fine-tuning of a transformer language model on either of these datasets can learn better answer retrieval and turn-by-turn question generation, fulfilling part of the requirements for CIS \cite{lewis2020retrieval}. 

\noindent Despite large-scale crowdsourced annotations, difficulty arises when a CIS agent is required to perform in the presence of a small size IS query. A search through multiple documents to generate logically inferential and semantically related questions for driving mixed-initiative conversations is required \cite{cho2021contrastive}. Also, the agent needs to utilize explicit knowledge to enrich IS queries and broaden the retriever's context to ask questions that would otherwise miss relevant information \cite{li2020enriching}. The datasets mentioned above lack the characteristics needed in training such a CIS agent. This motivates creating discourse datasets like CAsT-19 \cite{dalton2020cast}, used in this research for development and assessment of {\fontfamily{cmss}\selectfont ISEEQ}. Furthermore, there is no study close to {\fontfamily{cmss}\selectfont ISEEQ} that combines explicit knowledge, multi-passage retrieval, and question generation for CIS. Indirectly, {\fontfamily{cmss}\selectfont ISEEQ} gather specific attributes from retrieval augmented generation (RAG) model and retrieval augmented language model (REALM) from \cite{lewis2020retrieval} and \cite{guu2020realm}. However, it significantly improves upon them by (a) preserving the semantic and syntactic structure of the query, (b) use knowledge graphs for passage retrieval, and (c) maintain information flow in question generation. In our evaluations, we utilize RAG components (T5 and DPR) to measure accuracy and quality of ISQs. 

\section{Approach}
\label{sec:methods}
\begin{figure*}[!htbp]
    \centering
    \includegraphics[width=100mm, scale=0.50, trim=4.cm 5.6cm 6.0cm 3.8cm]{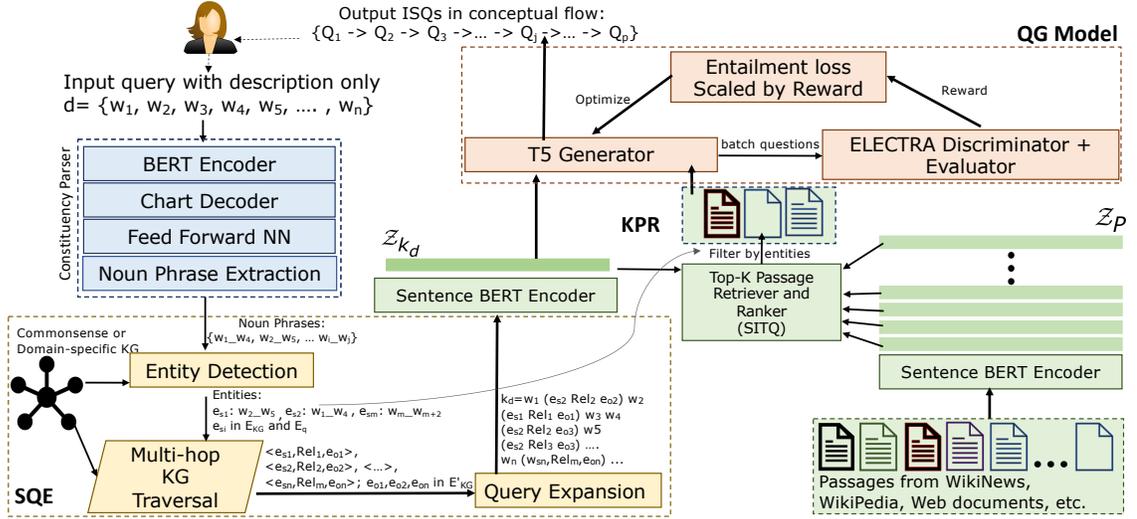}
  \caption{\footnotesize Overview of our approach. {\fontfamily{cmss}\selectfont ISEEQ} combines a BERT-based constituency parser, Semantic Query Expander (SQE), and Knowledge-aware Passage Retriever (KPR) to provide relevant context to a QG model for ISQ generations. The QG Model illustrates a structure of {\fontfamily{cmss}\selectfont ISEEQ} variants: {\fontfamily{cmss}\selectfont ISEEQ-RL} and {\fontfamily{cmss}\selectfont ISEEQ-ERL}.  We train {\fontfamily{cmss}\selectfont ISEEQ} in  generative-adversarial reinforcement learning setting that maximizes semantic relations and coherence while generating ISQs.}
  \label{fig:architecture_example}
\end{figure*}

\paragraph{Problem Definition:} 
Given a short query ($q = w_1, w_2, w_3,...., w_n$) on any topic (e.g., mental health, sports, politics and policy, location, etc.) automatically generate ISQs in a conceptual flow ($ISQ: Q_1, Q_2, Q_3, ..., Q_p$) to understand specificity in information needs of the user.

Our approach to address this problem, {\fontfamily{cmss}\selectfont ISEEQ}, is outlined in Figure \ref{fig:architecture_example}. We describe in detail the main components of {\fontfamily{cmss}\selectfont ISEEQ}: semantic query expander (SQE), knowledge-aware passage retriever (KPR) and generative-adversarial \textbf{R}einforcement \textbf{L}earning-based question generator ({\fontfamily{cmss}\selectfont ISEEQ-RL}) with \textbf{E}ntailment constraints ({\fontfamily{cmss}\selectfont ISEEQ-ERL}). Inputs to {\fontfamily{cmss}\selectfont ISEEQ} are IS queries described in natural language. For instance, an IS query can be described with \textbf{\texttt{Titles and Descriptions (T \& D)}} (such as in CAsT-19 dataset), \textbf{\texttt{Descriptions only (D only)}} (such as in QAMR and QADiscourse datasets), \textbf{\texttt{Topics and Aspects (Tp \& Asp)}} (such as in Facebook Curiosity discourse dataset), and others.

\textbf{SQE}: We expand the possibly short user input queries with the help of  ConceptNet Commonsense Knowledge Graph (CNetKG)\cite{speer2017conceptnet}. We first extract the entity set $\mathbf{E}_d$ in a user query description \texttt{d} using CNetKG. For this, we use the pre-trained self-attentive encoder-decoder-based  constituency parser \cite{kitaev2018constituency} with BERT as the encoder for consistency in {\fontfamily{cmss}\selectfont ISEEQ}. The parser is conditioned to extract noun phrases that capture candidate entities defining an IS query. If the phrases have mentions in the CNetKG they are termed as entities~\footnote{From here onwards we only use the term Entities, presuming check through exact match is performed using CNetKG}. Then a multi-hop triple (subject-entity, relation, object-entity) extraction over CNetKG is performed using depth first search on entity set $\mathbf{E}_d$. Triples of the form $<e_{d}, Rel_{i}, e_{x}>$ and $<e_y, Rel_{j}, e_{d}>$ are extracted where $e_{d}\in$ $\mathbf{E}_d$. We keep only those triples where $e_{d}$ ($\in$ $\mathbf{E}_d$) appears as the subject-entity.  We use this heuristic (1) to minimize noise and (2) gather more direct information about entities in $\mathbf{E}_d$. Finally, we contextualize \texttt{d} by injecting extracted triples to get $k_d$, a knowledge augmented query.

 Take for example \textbf{\texttt{D only}} IS query \texttt{d} ($\in \mathbf{D}$), ``Want to consider career options from becoming a physician's assistant vs a nurse''. The extracted entity set $\mathbf{E}_d$ for \texttt{d} is \{career, career\_options, physician, physician\_assistant, nurse\}. Then, the extracted triples for this entity set are $<$career\_options, isrelatedto, career\_choice$>$, $<$career\_options, isrelatedto, profession$>$, $<$physician\_assistant, is\_a, PA$>$, $<$ physician, is\_a, medical doctor$>$, [...], $<$nurse, is\_a, psychiatric\_nurse$>$, $<$nurse, is\_a, licensed\_practical\_nurse$>$, $<$nurse, is\_a, nurse\_practitioner$>$, [...]. The knowledge augmented $k_d$ is  ``Want to consider career options~\texttt{career\_options is related to career\_choice, profession} from becoming a physician's assistant \texttt{physician\_assistant is\_a PA medical doctor, [...]} vs a nurse \texttt{nurse is\_a psychiatric\_nurse, licensed\_practical\_nurse, [...] }''.
 Next, we pass this into KPR. The set \{$k_d\},~\forall \texttt{d} \in \mathbf{D}$ is denoted by $\mathbf{K_D}$ used by QG model in {\fontfamily{cmss}\selectfont ISEEQ}
 
\textbf{KPR:} Given the knowledge augmented query $k_d$, KPR retrieve passages from a set $\mathbf{P}$ and rank to get top-K passages $\mathbf{P}_{\mbox{top-K}}$. For this purpose, we make following specific improvements in the Dense Passage Retriever (DPR) described in \cite{lewis2020retrieval}: 
(1) Sentence-BERT encoder for the passages $p \in \mathbf{P}$ and $k_d$. We create dense encodings of $p \in \mathbf{P}$ using Sentence-BERT, which is represented as $\mathcal{Z}_{p}$ \cite{reimers2019sentence}. Likewise, encoding of $k_d$ is represented as $\mathcal{Z}_{k_d}$. (2) Incorporate SITQ (Simple locality sensitive hashing (Simple-LSH) and Iterative Quantization) algorithm to pick top-K passages ($\mathbf{P}_{\mbox{top-K}}$) by using a normalized entity score (NES). SITQ is a fast approximate search algorithm over MIPS to retrieve and rank passages. It can be formalized as $Score(\mathbf{P}_{\mbox{top-K}}|k_d)$ where,

\[
Score(\mathbf{P}_{\mbox{top-K}}|k_d) \propto \{\mbox{WMD}(\mathcal{Z}_{k_d}^T\mathcal{Z}_p)\}_{p \in \mathbf{P}}\]
\[\mathcal{Z}_{k_d} = \mbox{S-BERT}(k_d);
\mathcal{Z}_p= \mbox{S-BERT}(p);
\]
SITQ converts dense encodings into low-rank vectors and calculates the semantic similarity between the input query and passage using word mover distance (WMD) \cite{kusner2015word}. $\mathbf{P}_{\mbox{top-K}}$ from SITQ is re-ranked by NES, calculated\footnote{an entity occurring multiple times in $p$ is counted once} for each $p \in \mathbf{P}_{\mbox{top-K}}$ as $\frac{\sum_{e_j \in k_d} \{\mathbb{I}(e_j = w)\}_{w \in p}}{|k_d|}$ and arrange in descending order. $\mathbf{P}_{\mbox{top-K}}$ consists of $K$ passages with NES $>$80\%. Execution of KPR is iterative and stopped when each query in the train set has at least one passage for generating ISQs.

We tested retrieving efficiency of KPR using encoding of $e_d$ denoted by $\mathcal{Z}_{e_d}$ 
and using the encoding of $k_d$ denoted by $\mathcal{Z}_{k_d}$ as inputs to KPR.
Measurements were recorded using Hit Rate (HR) @ 10 and 20 retrieved passages. Mean Average Precision (MAP) is calculated with respect to ground truth questions in QAMR. There are two components in MAP: (a) \textit{Relevance} of the retrieved passage in generating questions that have $>$70\% cosine similarity with ground truth; (b) Normalize \textit{Relevance} by the number of ground truth questions per input query. To get MAP, we multiply (a) and (b) and take mean over all the input queries. We computed MAP by setting $K=20$ retrieved passages due to the good confidence from hit rate (a hyperparameter). KPR outperformed the comparable baselines on the QAMR Wikinews dataset and Table \ref{tab:retriever} shows that SQE improves the retrieval process\footnote{KPR($\mathcal{Z}_{e_d}$) \& KPR($\mathcal{Z}_{k_d}$) is executed for each CAsT-19 query.} A set of $\mathbf{P}_{\mbox{top-K}}$ for $\mathbf{K_D}$ is denoted by $\{\mathbf{P}_{\mbox{top-K}}\}_{k_d}, ~ k_d \in \mathbf{K_D}$.

\begin{table}[!htbp]
{\fontfamily{cmss}\selectfont
\footnotesize 
    \begin{tabular}{p{4.7cm}p{.7cm}p{.7cm}p{0.7cm}}
        \toprule[1.5pt]
         \textbf{Retrievers} & \textbf{HR@10} & \textbf{HR@20} & \textbf{MAP} \\ \midrule[1pt]
         TF-IDF + ECE {\cite{clark2019electra}} & 0.31 & 0.45 & 0.16 \\
         BM25 + ECE*  & 0.38 & 0.49 & 0.23 \\
         DPR {\cite{karpukhin2020dense}} & 0.44 & 0.61 & 0.31 \\ \midrule
         KPR($\mathcal{Z}_{e_d}$) & 0.47 & 0.66 & 0.35 \\
         \rowcolor{Gray}KPR($\mathcal{Z}_{k_d}$) &0.49 & 0.70 & 0.38 \\ \bottomrule[1.5pt]
    \end{tabular}}
    \caption{\footnotesize Evaluating retrievers. ECE: Electra Cross Encoder, (*): variant of (Clark et al. 2019), DPR: Dense Passage Retrieval.}    
    \label{tab:retriever}
\end{table}

\noindent \textbf{QG Model:} {\fontfamily{cmss}\selectfont ISEEQ} leverages $\mathbf{K_D}$ and $\{\mathbf{P}_{\mbox{top-K}}\}_{k_d}$ to learn QG in generative-adversarial setting guided by a reward function. {\fontfamily{cmss}\selectfont ISEEQ-RL} contains T5-base as generator and Electra-base as discriminator to learn to generate IS-type questions. {\fontfamily{cmss}\selectfont ISEEQ} use the reward function to learn to selectively preserve terms from the IS query versus introducing diversity. Also, reward function prevent {\fontfamily{cmss}\selectfont ISEEQ} from generating ISQs that are loose in context or redundant. 

\noindent \textbf{Reward Function:} Let $q^n_i$ be the $i^{th}$ question in the ground truth questions $Q$ having $n$ tokens and let $\hat{q}^m_i$ be the $i^{th}$ question in the list of generated questions, $\hat{Q}$ having $m$ tokens. We create BERT encodings for each of the $n$ and $m$ words in the question vectors. The reward ($R_{i}$) in {\fontfamily{cmss}\selectfont ISEEQ-RL} and {\fontfamily{cmss}\selectfont ISEEQ-ERL} is defined as:
\begin{equation}
\footnotesize
\label{eq:reward}
    \alpha\bigg[\frac{LCS(\hat{q}_{i}^{m},q_{i}^{n})}{|\hat{q}_{i}^{m}|}\bigg]+(1-\alpha)\bigg[\sum_{\hat{w_{ij}}\in\hat{q}_{i}^{m}}\max_{w_{ik}\in q_{i}^{n}}\mbox{WMD}(\hat{w_{ij}}^{T}w_{ik})\bigg]
\end{equation}
\noindent where $\alpha[*]$ is a normalized longest common subsequence (LCS) score that capture word order and make {\fontfamily{cmss}\selectfont ISEEQ-RL} learn to copy in some very complex IS-type queries. $(1-\alpha)[*]$ uses WMD to account for semantic similarity and compositional diversity. For a $q^n_i$ = ``What is the average starting salary in the UK?'', $(1-\alpha)[*]$ generates $\hat{q}^m_i$=``What is the average earnings of nurse in UK?''

\noindent \textbf{Loss Function in {\fontfamily{cmss}\selectfont ISEEQ-RL}:} We revise cross entropy (CE) loss for training {\fontfamily{cmss}\selectfont ISEEQ} by scaling with the reward function because each $k_d \in \mathbf{K_D}$ are not only short but they also vary by context. Corresponding to each $k_d$, there are $b$ ground truth questions $q_{1:b}$ and thus we normalize the revised CE loss by a factor of $b$. Formally, we define our CE loss in  {\fontfamily{cmss}\selectfont ISEEQ-RL}, $\mathcal{L}(\hat{q}_{1:b}|q_{1:b},\theta)=$ 
\begin{equation}
\label{eq:ce}
   \frac{-\sum_{i=1}^b R_i\cdot\mathbb{I}(q^n_i=\hat{q}^m_i)\cdot\mathbf{log}Pr(\hat{q}^m_i|\theta)}{b}
\end{equation}
\noindent where $\mathbb{I}(q^n_i=\hat{q}^m_i)$ is an indicator function counting word indices in $q^n_i$ that match word indices in $\hat{q}^m_i$. The CE loss over $\mathbf{K_D}$ in a discourse dataset is $\mathcal{L}(\hat{Q}|Q,\Theta)_t$, recorded after $t^{th}$ epoch. Formally $\mathcal{L}(\hat{Q}|Q,\Theta)_t$ =
\begin{equation}
\label{eq:tune}
\gamma \mathcal{L}(\hat{Q}|Q,\Theta)_{t-1} + (1-\gamma)\mathcal{L}(\hat{q}_{1:b}|q_{1:b},\theta)
\end{equation}
Theoretically, {\fontfamily{cmss}\selectfont ISEEQ-RL} addresses RQ1, but weakly mandates conceptual flow while generating ISQs. Thus, it does not address RQ2. 

\noindent \textbf{Loss Function in {\fontfamily{cmss}\selectfont ISEEQ-ERL}:}
For instance, given \texttt{d2}($\in \mathbf{D}$): ``Bothered by feeling down or depressed'' (shown in Figure \ref{fig:intro_figure}), {\fontfamily{cmss}\selectfont ISEEQ-RL} generations are: ($\hat{q}_1$): What is the reason for the depression, hopelessness? and ($\hat{q}_2$) What is the frequency of you feeling down and depressed? Whereas, {\fontfamily{cmss}\selectfont ISEEQ-ERL} would re-order placing ($\hat{q}_2$) before ($\hat{q}_1$) for conceptual flow. To develop {\fontfamily{cmss}\selectfont ISEEQ-ERL}, we redefine the loss function in {\fontfamily{cmss}\selectfont ISEEQ-RL} by introducing principles of entailment as in NLI \cite{tarunesh2021trusting}\cite{gao2020discern}\footnote{We use RoBERTa pre-trained on Stanford NLI dataset to measure semantic relations and coherence between a pair of generated questions}. Consider $\hat{q}^m_{i|next}$ to be the next generated question after $\hat{q}^m_{i}$. We condition equation \ref{eq:ce} on $y_{max} = \argmax_{Y} \mbox{RoBERTa}(\hat{q}^m_i, \hat{q}^m_{i|next})$, where $Y \in$ \{neutral, contradiction, entailment\} and $Pr(y_{max}) = \max_{Y}\mbox{RoBERTa}(\hat{q}^m_i, \hat{q}^m_{i|next})$. Formally, $\mathcal{L}(\hat{q}_{1:b}|q_{1:b},\theta)$ in {\fontfamily{cmss}\selectfont ISEEQ-ERL}is: 

\begin{algorithmic}
\label{eq:4}
  \IF{$y_{max} == \mbox{entailment}$}
    \STATE  $ \mbox{CE} - Pr(y_{max})$
  \ELSE
    \STATE $ \mbox{RCE}-(1-Pr(y_{max}))$
  \ENDIF
\end{algorithmic}
\mbox{RCE}= $-\frac{\sum_{i=1}^b R_i(1-\mathbb{I}(q_i=\hat{q}_i))Pr(\hat{q}_i|\theta)}{b}$ \newline Reverse Cross Entropy(RCE) complements CE (Equation \ref{eq:ce}) by checking $\hat{q}^m_{i|next}$ is semantically related and coherent to $\hat{q}^m_i$. Tuning of the loss after an epoch follows Equation \ref{eq:tune}. 

\section{Datasets} 
\label{sec:datasets}
\begin{table}[!ht]
\footnotesize
{\fontfamily{cmss}\selectfont
    \begin{tabular}{p{1.3cm}p{1.2cm}p{1.3cm}p{0.8cm}c}
        \toprule[1.5pt]
         \textbf{Dataset} & \multicolumn{2}{c}{\textbf{\#Queries}(\textbf{Q/Q})}  & \textbf{CNetKG Triples} & \textbf{Source}\\ \cmidrule{2-3}
         & Train & Test &  & \\
         QAD & 125 (25) & 33 (25) & 38.5\% & WikiP, WikiN \\
         QAMR & 395 (63) & 39 (68) & 35.5\% & WikiN \\ 
         FBC & 8489 (6) & 2729 (8) & 50\% & Geo-WikiP \\
         CAsT-19 & 30 (9) & 50 (10)  & 57\% & MS-MARCO \\ \bottomrule[1.5pt]
    \end{tabular}}
    \caption{\footnotesize Dataset description. \textbf{Q/Q:} Questions per Query, CNetKG Triples: \% of noun/verb phrases identified in CNetKG.}
    \label{tab:datasets}
\end{table}

\begin{table*}[t]
\scalebox{0.8}{
{\fontfamily{cmss}\selectfont
\begin{tabular}{p{3.2cm}p{1.7cm}p{0.7cm}p{0.7cm}p{0.7cm}p{0.5cm}p{0.8cm}|p{0.7cm}p{0.7cm}p{0.7cm}p{0.5cm}p{0.8cm}|p{0.7cm}p{0.7cm}p{0.7cm}p{0.5cm}p{0.8cm}}
\toprule[1.5pt]
\textbf{Methods} & \textbf{SQE} & \multicolumn{5}{c}{\textbf{QAD}} & \multicolumn{5}{c}{\textbf{QAMR}} & \multicolumn{5}{c}{\textbf{FBC}} \\ 
& & R-L & BRT & BScore & SR & LC(\%) & R-L & BRT & BScore & SR & LC(\%) & R-L & BRT & BScore & SR & LC(\%) \\ \toprule[1.5pt]
\multirow{3}{*}{T5-FT WikiPassageQA} & - & 0.37 & 0.43 & 0.16 & 0.17 & 10.0 & 0.19 & 0.51 & 0.38 & 0.36 & 17.0 
& 0.65 & 0.78 & 0.54 & 0.51 & 47.3 \\
& +\textcolor{OliveGreen}{Entities} & 0.39 & 0.45 & 0.16 & 0.17 & 10.0 & 0.20 & 0.53 & 0.38 & 0.36 & 17.5 
& 0.65 & 0.78 & 0.54 & 0.52 & 47.4 \\  
 & +\textcolor{blue}{Triples}& 0.41 & 0.46 & 0.16 & 0.18 & 11.0 & 0.20 & 0.53 & 0.39 & 0.37 & 17.8 &  
0.65 & 0.78 & 0.55  & 0.52 & 47.3 \\ \cmidrule{2-17}
\multirow{3}{*}{T5-FT SQUAD} & - & 0.44 & 0.54 & 0.20 & 0.19& 13.0 & 0.40 & 0.66 & 0.46 & 0.58 & 21.0 
& 0.70 & 0.83 & 0.62 & 0.67 & 65.1\\ 
& +\textcolor{OliveGreen}{Entities} & 0.45 & 0.56 & 0.22 & 0.19 & 13.5 & 0.40 & 0.68 & 0.47 & 0.59 & 22.7 
& 0.71 & 0.84 & 0.63 & 0.69 & 65.8 \\ 
 & +\textcolor{blue}{Triples} & 0.45 & 0.58 & 0.22 & 0.20 & 13.8 & 0.43 & 0.69 & 0.47 & 0.59 & 22.6  
& 0.70 & 0.84 & 0.64 & 0.69 & 65.8 \\ \cmidrule{2-17}
\multirow{3}{*}{T5-FT CANARD} & - & 0.47 & 0.54 & 0.23 & 0.19 & 17.1 & 0.41 & 0.64 & 0.53 & 0.58 & 22.6 
& 0.73 & 0.84 & 0.63 & 0.67 & 66.2 \\
& +\textcolor{OliveGreen}{Entities} & 0.48 & 0.55 & 0.25 & 0.20 & 17.5 & 0.44 & 0.67 & 0.62 & 0.61 & 23.5 
& 0.74 & 0.84 & 0.65 & 0.69 & 66.5 \\
& +\textcolor{blue}{Triples} & 0.51 & 0.57 & 0.26 & 0.21 & 18.3 & 0.49 & 0.68 & 0.66 & 0.61 & 24.3 
& 0.74 & 0.85 & 0.65 & 0.70 & 68.2 \\ \cmidrule{2-17}
\multirow{3}{*}{ProphetNet-FT SQUAD} &-& 0.31 & 0.44 & 0.14 & 0.17 & 12.2 & 0.35 & 0.59 & 0.38 & 0.36 & 21.5 
& 0.63 & 0.78 & 0.53 & 0.67 & 63.2 \\
 & +\textcolor{OliveGreen}{Entities} & 0.31 & 0.44 & 0.14 & 0.17 & 12.7 & 0.37 & 0.60 & 0.41 & 0.37 & 22.1 
& 0.65 & 0.78 & 0.54 & 0.67 & 63.3 \\
& +\textcolor{blue}{Triples} & 0.34 & 0.45 & 0.15 & 0.18 & 13.0 & 0.37 & 0.61 & 0.43 & 0.37 & 22.3 
& 0.65 & 0.79 & 0.56 & 0.69 & 64.0 \\ \bottomrule[1.2pt]
\multirow{3}{*}{{\fontfamily{cmss}\selectfont ISEEQ-RL}} & - & 0.57 & 0.72 & 0.40 & 0.22 & 20.0 & 0.50 & 0.75 & 0.67 & 0.64 & 29.4 & 0.71 & 0.84 &0.62 & 0.69 & 68.2  \\
& +\textcolor{OliveGreen}{Entities} & 0.64 & 0.72 & 0.41 & 0.23 & 22.0 & 0.52 & 0.77 & 0.68 & 0.64 & 33.1 & 0.72 & 0.85 & 0.63 & 0.71 & 69.8 \\
& +\textcolor{blue}{Triples} & \cellcolor{Gray}0.65 & \cellcolor{Gray}0.74 & \cellcolor{Gray}0.45 & \cellcolor{Gray}0.25 & \cellcolor{Gray}22.0 & \cellcolor{Gray}0.53 & \cellcolor{Gray}0.78 & \cellcolor{Gray}0.71 & \cellcolor{Gray}0.65 & \cellcolor{Gray}34.7 & \cellcolor{Gray}0.74 & \cellcolor{Gray}0.87 & \cellcolor{Gray}0.63 & \cellcolor{Gray}0.73 & \cellcolor{Gray}71.8 \\ \cmidrule{2-17}
\multirow{3}{*}{{\fontfamily{cmss}\selectfont ISEEQ-ERL}} & - & 0.60 & 0.76 & 0.44 & 0.26 & 24.5 & 0.55 & 0.81 & 0.72 & 0.68 & 36.1 
&0.74 & 0.85 & 0.64 & 0.76 & 78.2 \\
& +\textcolor{OliveGreen}{Entities} & 0.65 & 0.78 & 0.47 & 0.27 & 25.2 & 0.55 & 0.82 & 0.74 & 0.68 & 36.3 
& 0.77 & 0.88 & 0.66 & 0.76 & 78.3 \\
& +\textcolor{blue}{Triples} & \cellcolor{Gray}\textbf{0.67} & \cellcolor{Gray}\textbf{0.79} & \cellcolor{Gray}\textbf{0.50} & \cellcolor{Gray}\textbf{0.27} & \cellcolor{Gray}\textbf{25.7} & \cellcolor{Gray}\textbf{0.57} & \cellcolor{Gray}\textbf{0.83} & \cellcolor{Gray}\textbf{0.77} & \cellcolor{Gray}\textbf{0.68} & \cellcolor{Gray}\textbf{37.0} & \cellcolor{Gray}\textbf{0.79} & \cellcolor{Gray}\textbf{0.89} & \cellcolor{Gray}\textbf{0.66} & \cellcolor{Gray}\textbf{0.78} & \cellcolor{Gray}\textbf{79.4}\\
\bottomrule[1.5pt]
\end{tabular}}}
\caption{\footnotesize Scores on test set of datasets. In comparison to T5-FT CANARD, a competitive baseline, {\fontfamily{cmss}\selectfont ISEEQ-ERL} generated better questions across three datasets (30\%$\uparrow$ in QADiscourse, 7\%$\uparrow$ in QAMR, and 5\%$\uparrow$ in FB Curiosity). For fine-tuning we used SQUADv2.0.}
\label{tab:major_results}
\end{table*}

We evaluate {\fontfamily{cmss}\selectfont ISEEQ-RL} and {\fontfamily{cmss}\selectfont ISEEQ-ERL} on a wide range of open-domain knowledge-intensive datasets. Their statistics are shown in Table \ref{tab:datasets}. The datasets exhibit following properties: (1) existence of semantic relations between questions, (2) logical coherence between questions, and (3) diverse context, that is, queries cover wider domains, such as health, sports, history, geography. Fundamentally, these datasets support the assessment of RQ1, RQ2, and RQ3.  

QADiscourse (QAD) \cite{pyatkin2020qadiscourse} dataset tests the ability of {\fontfamily{cmss}\selectfont ISEEQ} to generate questions that have logical coherence. The sources of queries are Wikinews (WikiN) and Wikipedia (WikiP) that consist of 8.7 Million passages. Question Answer Meaning Representation (QAMR) \cite{michael2018crowdsourcing} dataset tests the ability of {\fontfamily{cmss}\selectfont ISEEQ} to generate questions with semantic relations between them. The source for creating IS queries is Wikinews, which consist of 3.4 Million passages. Both QAD and QAMR consist of \textbf{\texttt{D only}} IS queries. 
Facebook Curiosity (FBC) \cite{rodriguez2020information} is another dataset that challenges {\fontfamily{cmss}\selectfont ISEEQ} to have both semantic relations and logical coherence. This is because queries are described in the form of \textbf{\texttt{Tp \& Asp}}. The source for IS queries is Wikipedia having 3.3 Million geographical passages. Even though the questions in the dataset have logical coherence, they are relatively less diverse than QAMR and QAD. Conversational Assistance Track (CAsT-19) \cite{dalton2020cast} is the most challenging one for {\fontfamily{cmss}\selectfont ISEEQ} because of size, diversity in context, large number of passages, and IS queries are not annotated with passages. In CAsT-19, IS queries are provided with \textbf{\texttt{T \& D}}.

\textbf{Adapting Datasets:} Each dataset, except CAsT-19, has a query, a set of ISQs, and a relevant passage. For fairness in evaluation, we exclude the passages in the datasets; instead, we retrieve them from the sources using KPR. We also perform coreference resolution over ISQs using  NeuralCoref  to increase entity mentions \cite{clark2016improving}. For example, a question in CAsT-19 ``What are the educational requirements required to become one?'' is reformulated to ``What are the educational requirements required to become a physician's assistant?''. 

\section{Evaluation and Results}
\label{sec:eval_metrics}
{\fontfamily{cmss}\selectfont ISEEQ-RL or ISEEQ-ERL} generator uses top-p (nucleus) sampling \footnote{Top-p or Top-K sampling either works in {\fontfamily{cmss}\selectfont ISEEQ}} with sum probability of generations equaling to 0.92, a hyperparameter that sufficiently removes the possibility of redundant QG \cite{holtzman2019curious}. We evaluate {\fontfamily{cmss}\selectfont ISEEQ} generations using Rouge-L (R-L), BERTScore (BScore) \cite{zhang2019bertscore}, and BLEURT (BRT) \cite{sellam2020bleurt} that measure preservation of syntactic context, semantics, and legibility of generated question to human understanding, respectively. For conceptual flow in question generation, we define ``semantic relations'' (SR) and ``logical coherence'' (LC) metrics. To calculate SR or LC, we pair $\hat{Q}_{1:p}$ generated questions with $Q$. SR in the generations is computed across all pairs using RoBERTa pre-trained on semantic similarity tasks\footnote{https://huggingface.co/textattack/roberta-base-STS-B}. LC between $Q$ and $\hat{Q}_{1:p}$ is computed from counting the labels predicted as ``entailment'' by RoBERTa pre-trained on SNLI dataset\footnote{https://paperswithcode.com/lib/allennlp/roberta-snli}. 

\begin{table}[!ht]
\footnotesize
{\fontfamily{cmss}\selectfont
    \begin{tabular}{p{1.cm}|p{1.2cm}|p{1.5cm}|p{1.5cm}|p{1.4cm}}
    \toprule[1.5pt]
    \multirow{2}{*}{\textbf{{{\fontfamily{cmss}\selectfont ISEEQ}}}} & \multirow{2}{*}{\textbf{Encoding}} & \textbf{QAD} & \textbf{QAMR} & \textbf{FBC} \\ \cmidrule{3-5}
   Models & & \multicolumn{3}{c}{SR/LC(\%)} \\ \midrule[1pt]
    \multirow{2}{*}{RL} & $P_{1:K}$ & 0.21/ 21.3 & 0.63/ 28.0 & 0.73/ 71.6 \\ \cmidrule{2-5}              
       & $P_{1:K}$+$k_d$ & 0.25/ 22.0 & 0.65/ 34.7 & 0.73/ 71.8\\ \midrule[1pt]
    \multirow{2}{*}{ERL} & $P_{1:K}$ & 0.26/ 24.6 & 0.68/ 36.3 & 0.77/ 78.2 \\\cmidrule{2-5}
    & $P_{1:K}$+$k_d$ & 0.27/ 25.7 & 0.68/ 37.0 & 0.78/ 79.4 \\ \bottomrule[1.5pt]
    \end{tabular}}
    \caption{\footnotesize Ablation study to show importance of SQE in improving semantic relations and logical coherence of generated ISQs. Encodings of $\mathbf{P}_{1:K}$ and knowledge-augmented query ($k_d$) from SQE, were concatenated. Concatenation performed for each $p \in \mathbf{P}_{1:K}$.}
    \label{tab:ablation}
\end{table}

\textbf{Baselines:} Since there exists no system to automatically generate ISQs, we considered transformer language models fine-tuned (TLMs-FT) on open domain datasets used for reading comprehension, and complex non-factoid answer retrieval as baselines. Specifically, T5 model fine-tuned (T5-FT) on WikipassageQA \cite{cohen2018wikipassageqa}, SQUAD \cite{raffel2019exploring}, and CANARD \cite{lin2020conversational}, and ProphetNet\cite{qi2020prophetnet} fine-tuned on SQUADv2.0 are comparable baselines.

We substantiate our claims in RQ1, RQ2, and RQ3 by highlighting: (1) Multiple passage-based QG yields better results over single gold passage QG used in TLMs-FT (Table \ref{tab:major_results}); (2) Knowledge-infusion through SQE significantly advance the process of QG (Table \ref{tab:ablation}); (3) Pressing on conceptual flow in {\fontfamily{cmss}\selectfont ISEEQ-ERL} improve SR and LC in generations. Evidence from 12 human evaluations support our quantitative findings (Table \ref{tab:human_eval}); (4) We investigate the potential of {\fontfamily{cmss}\selectfont ISEEQ-ERL} in minimizing crowd workers for IS dataset creation through cross-domain experiments (Table \ref{tab:domain_adapt}).

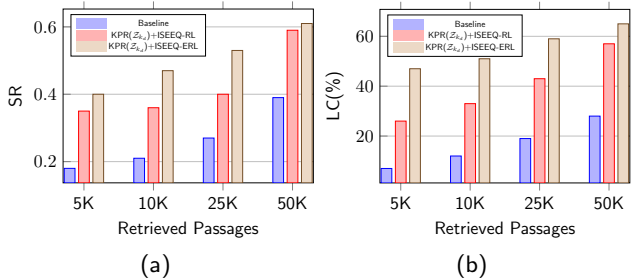
\begin{figure}[t]
{\fontfamily{cmss}\selectfont
\begin{minipage}[h]{1.0\linewidth}
\centering
\subfloat[][]{\resizebox{0.5\textwidth}{!}{
\begin{tikzpicture}
\begin{axis}[legend style={nodes={scale=0.5, transform shape}}, width=6.5cm, height=5.cm,
    ybar,
    bar width=6pt,
    xlabel={Retrieved Passages},
    ylabel={SR},
    xtick=data,
    xticklabels from table={\datatable}{A},
    ymajorgrids,
    legend pos=north west,
    legend image code/.code={%
                    \draw[#1, draw=none] (0cm,-0.1cm) rectangle (0.6cm,0.1cm);
                }
             ]
    \addplot table [x expr=\coordindex, y= B]{\datatable};
    \addplot table [x expr=\coordindex, y= C]{\datatable};
    \addplot table [x expr=\coordindex, y= D]{\datatable};
    \legend{Baseline, KPR($\mathcal{Z}_{k_d}$)+{\fontfamily{cmss}\selectfont ISEEQ-RL}, KPR($\mathcal{Z}_{k_d}$)+{\fontfamily{cmss}\selectfont ISEEQ-ERL}}
\end{axis}
\end{tikzpicture}}}
\subfloat[][]{\resizebox{0.5\textwidth}{!}{
\begin{tikzpicture}
\begin{axis}[legend style={nodes={scale=0.5, transform shape}}, width=6.5cm, height=5.cm,
    ybar,
    bar width=6pt,
    xlabel={Retrieved Passages},
    ylabel={LC(\%)},
    xtick=data,
    xticklabels from table={\datatables}{A},
    ymajorgrids,
    legend pos=north west,
    legend image code/.code={%
                    \draw[#1, draw=none] (0cm,-0.1cm) rectangle (0.6cm,0.1cm);
                }
             ]
    \addplot table [x expr=\coordindex, y= B]{\datatables};
    \addplot table [x expr=\coordindex, y= C]{\datatables};
    \addplot table [x expr=\coordindex, y= D]{\datatables};
    \legend{Baseline, KPR($\mathcal{Z}_{k_d}$)+{\fontfamily{cmss}\selectfont ISEEQ-RL}, KPR($\mathcal{Z}_{k_d}$)+{\fontfamily{cmss}\selectfont ISEEQ-ERL}}
\end{axis}
\end{tikzpicture}}}
\end{minipage}}
\caption{\footnotesize Performance improvement of {\fontfamily{cmss}\selectfont ISEEQ-ERL} over {\fontfamily{cmss}\selectfont ISEEQ-RL} and Baseline: T5-FT CANARD concerning SR and LC in generated ISQs. Performed on CAsT-19 with \textit{unannotated} passages.}
\label{fig:cast19_results}
\end{figure}

\textbf{Performance of {\fontfamily{cmss}\selectfont ISEEQ-RL} and {\fontfamily{cmss}\selectfont ISEEQ-ERL} :} Datasets used in this research were designed for a CIS system to obtain the capability of multiple contextual passage retrieval and diverse ISQ generations. The process of creating such datasets requires crowd workers to take the role of a CIS system responsible for creating questions and evaluators to see whether questions match the information needs of IS queries. Implicitly, the process embed crowd workers' curiosity-driven search to read multiple passages for generating ISQs. Baselines on employed datasets use single passage QG, with much of the efforts focusing on improving QG. Whereas {\fontfamily{cmss}\selectfont ISEEQ} generation enjoys the success from the connection of SQE, KPR, and novel QG model over baselines in CIS (see Table \ref{tab:major_results}). With SQE, {\fontfamily{cmss}\selectfont ISEEQ} achieved \mbox{2-6\%} across all datasets.
The knowledge-infusion in {\fontfamily{cmss}\selectfont ISEEQ} through SQE has shown to be powerful for baselines as well. Table \ref{tab:major_results} records \mbox{3-10\%, 3-10\%, and 1-3\%} performance gains of the baselines on QAD, QAMR, and FBC across five evaluation metrics, respectively. SQE allows baselines to semantically widen their search over the gold passages in datasets to generate diverse questions that match better with ground truth. Differently, {\fontfamily{cmss}\selectfont ISEEQ-RL} generations benefit from dynamic meta-information retrieval from multiple passages yielding hike of \mbox{20-35\%, 6-13\%, 3-10\%} on QAD, QAMR, and FBC, respectively, across five evaluation metrics. Especially, QG in CAsT-19 and FBC datasets advance because of KPR in  {\fontfamily{cmss}\selectfont ISEEQ-RL} and {\fontfamily{cmss}\selectfont ISEEQ-ERL} (see Figure \ref{fig:cast19_results}). Most of the CAsT-19 and FBC queries required multiple passages to construct legible questions. For instance, an IS query : ``Enquiry about History, Economy, and Sports in Hyderabad'' {\fontfamily{cmss}\selectfont ISEEQ} retrieved following three passages: ``History\_Hyderabad'', ``Economy\_Hyderabad'', and ``Sports\_Hyderabad'' which were missing in the set of passages in FBC. Thus, TLM-FT baselines find it hard to construct legible ISQs using a single passage. Furthermore, {\fontfamily{cmss}\selectfont ISEEQ-ERL} advances the quality of ISQs over {\fontfamily{cmss}\selectfont ISEEQ-RL} by 7-19\%, 4-7\%, and 5-6\% in QAD, QAMR, and FBC (see Table \ref{tab:ablation}) datasets.  This is because QAD and FBC questions require the QG model to emphasize conceptual flow.

\begin{table}[!ht]
\footnotesize
{\fontfamily{cmss}\selectfont
\begin{tabular}{c|p{0.7cm}|p{0.7cm}|p{0.7cm}|p{0.7cm}|p{0.7cm}|p{0.7cm}}
        \toprule[1.5pt]
         \textbf{Ret.Pass.} & \multicolumn{2}{c}{\textbf{DPR}} & \multicolumn{2}{c}{\textbf{KPR($\mathcal{Z}_{e_d}$)}} & \multicolumn{2}{c}{\textbf{KPR($\mathcal{Z}_{k_d}$)}}  \\ \cmidrule{2-7}
         & Train & Test & Train & Test & Train & Test \\ \midrule[1pt]
         5K & 71 & 123 & 99 & 278 & 157 & 275 \\
         10K & 96 & 133 & 154 & 301 & 194 & 316 \\
         25K & 139 & 133 & 235 & 329 & 236 & 363 \\
         50K & 173 & 144 & \cellcolor{Gray} \textbf{269} & 358 & \cellcolor{Gray}\textbf{269} & 402 \\ \bottomrule[1.5pt]
    \end{tabular}}
    \caption{\footnotesize Performance of KPR on MS-MARCO passages while retrieving atleast one passage per IS query in CAsT-19. 269 is the size of CAST-19 train set. KPR covered the train set but left 16\% of the IS queries in test set. }
    \label{tab:cast19_1}
\end{table}

Further, we examine the combined \textbf{performance of KPR} and  {\fontfamily{cmss}\selectfont ISEEQ-ERL} on CAsT-19 dataset. KPR retrieved $\sim$50K passages sufficient to generate questions for 269 IS queries\footnote{one query can have multiple passages}. Table \ref{tab:cast19_1} depicts KPR($\mathcal{Z}_{e_d}$) retrieval performance match KPR($\mathcal{Z}_{k_d}$), with later supported 72\% of queries in training set compare to 57\% by KPR($\mathcal{Z}_{e_d}$). Also, it outperforms DPR, which supported 30\% queries in train set (see Table \ref{tab:cast19_1}). In test time, KPR($\mathcal{Z}_{k_d}$) supported 84\% queries that were used to generate questions by {\fontfamily{cmss}\selectfont ISEEQ-ERL} and evaluated with ground truth for SR and LC (see Figure \ref{fig:cast19_results}). Apart from monotonic rise in SR and LC scores shown by {\fontfamily{cmss}\selectfont ISEEQ}, {\fontfamily{cmss}\selectfont ISEEQ-ERL} generations achieved better coherence than counterparts with 5K passages ( Figure \ref{fig:cast19_results} (c) \& (d)). We attribute the addition of entailment check and RCE for conceptual flow-based QG improvements. 
\textbf{Note:} Qualitative samples of ISQs generated by {\fontfamily{cmss}\selectfont ISEEQ-ERL}, {\fontfamily{cmss}\selectfont ISEEQ-RL} and Baseline (T5-FT CANARD) are provided \href{https://github.com/manasgaur/AAAI-22}{\textbf{here}} for comparison with ground truth ISQs.

\begin{table}[!ht]
\scriptsize
\scalebox{0.9}{
{\fontfamily{cmss}\selectfont
\begin{tabular}{p{1cm}p{1.5cm}p{1.5cm}p{1.5cm}p{1.5cm}}
\toprule[1.5pt]
Test $\rightarrow$ &  \textbf{QAD} & \textbf{QAMR}  & \textbf{FBC} & \textbf{CAsT19} \\ \cmidrule{2-5}
Train $\downarrow$ & \multicolumn{3}{c}{R-L/BRT/BScore/SR/LC(\%)} \\ \midrule[1pt] 
\textbf{QAD} & \cellcolor{Gray}\textbf{0.67/ 0.79/ 0.50/ 0.27/ 25.7} & \cellcolor{Blue}0.56/ 0.79/ 0.75/ 0.64/ 33.1 & 0.62/ 0.70/ 0.55/ 0.71/ 73.5 & 0.76/ 0.48/ 0.64/ 0.60/ 64.2\\
\textbf{QAMR} & 0.73/ 0.89/ 0.62/ 0.28/ 27.7 & \cellcolor{Gray}\textbf{0.57/ 0.83/ 0.77/ 0.68/ 37.0} & \cellcolor{Blue}0.74/ 0.89/ 0.67/ 0.75/ 77.8 & 0.67/ 0.41/ 0.57/ 0.57/ 58.6\\
\textbf{FBC} & 0.70/ 0.73/ 0.56/ 0.31/ 33.0 & 0.61/ 0.85/ 0.72/ 0.67/ 35.8 & \cellcolor{Gray}\textbf{0.79/ 0.89/ 0.66/ 0.78/ 79.4} & 0.75/ 0.37/ 0.76/ 0.67/ 66.5 \\
\textbf{CAsT-19} & 0.58/ 0.69/ 0.51/ 0.23/ 25.2 & \cellcolor{Blue}0.52/ 0.73/ 0.70/ 0.61/ 33.4 & 0.63/ 0.77/ 0.57/ 0.73/ 76.5 & \cellcolor{Gray}\textbf{0.74/ 0.48/ 0.68/ 0.61/ 65.0} \\ \bottomrule[1.5pt]
\end{tabular}}}
\caption{\footnotesize Transferability test scores using {\fontfamily{cmss}\selectfont ISEEQ-ERL} to answer RQ3. gray cell: {\fontfamily{cmss}\selectfont ISEEQ-ERL} trained and tested on same dataset. dark gray cell: shows acceptable cross-domain \{Train-Test\} pairs, where train size is smaller than test size.}
\label{tab:domain_adapt}
\end{table}

\textbf{Transferability Test for RQ3:} We examine the performance of {\fontfamily{cmss}\selectfont ISEEQ-ERL} in an environment where the train and test dataset belong to a different domain. For instance, QAMR is composed of IS queries from Wikinews, whereas FBC is composed of IS queries from geography category in Wikipedia. From experiments in Table \ref{tab:domain_adapt}, we make two deductions: (1) {\fontfamily{cmss}\selectfont ISEEQ-ERL} provided acceptable performance in generating ISQs for \{Train-Test\} pairs, where train size is smaller than test size: \{QAD-QAMR\} and \{QAMR-FBC\}.
(2) {\fontfamily{cmss}\selectfont ISEEQ-ERL} trained on a \emph{narrow domain dataset} (FBC) generated far better ISQs for IS queries in generic domain. The transferability test show {\fontfamily{cmss}\selectfont ISEEQ-ERL}'s ability to create new datasets for training and development of CIS systems.

\begin{table}[]
\footnotesize
{\fontfamily{cmss}\selectfont
\begin{tabular}{lllllp{1.7cm}}
\hline
\multirow{2}{*}{} & \multicolumn{3}{c}{Response: Mean (SD)}                                                                                                                                     & \multirow{2}{*}{\begin{tabular}[c]{@{}l@{}}F(2, 957)\\ (p-value)\end{tabular}} & \multirow{2}{*}{\begin{tabular}[c]{@{}l@{}}LSD post-hoc\\ (p \textless 0.05)\end{tabular}} \\ \cline{2-4}
                  & \multicolumn{1}{c}{S1}                                  & \multicolumn{1}{c}{S2}                                  & \multicolumn{1}{c}{S3}                                  &                                                                                &                                                                                            \\ \hline
G1                & \begin{tabular}[c]{@{}l@{}}3.756 \\ (1.14)\end{tabular} & \begin{tabular}[c]{@{}l@{}}3.759 \\ (1.06)\end{tabular} & \begin{tabular}[c]{@{}l@{}}3.518 \\ (1.08)\end{tabular} & \begin{tabular}[c]{@{}l@{}}5.05 \\ (6.5e-3)\end{tabular}                       & S1\textgreater{}S3, S2\textgreater{}S3                                                     \\ \hline
G2                & \begin{tabular}[c]{@{}l@{}}3.803 \\ (1.10)\end{tabular} & \begin{tabular}[c]{@{}l@{}}3.843 \\ (1.02)\end{tabular} & \begin{tabular}[c]{@{}l@{}}3.503 \\ (1.06)\end{tabular} & \begin{tabular}[c]{@{}l@{}}9.71 \\ (6.63e-5)\end{tabular}                      & S1\textgreater{}S3, S2\textgreater{}S3                                                     \\ \hline
\end{tabular}}
\caption{\footnotesize Assessment of human evaluation. G1: ISQs are diverse in context and non-redundant. G2: ISQs are logically coherent and share semantic relations. $>$: difference is statistically significant. SD: Standard Deviation. S1, S2, and S3 are ground truth, {\fontfamily{cmss}\selectfont ISEEQ-ERL}, and T5-FT CANARD, respectively.}
\label{tab:human_eval}
\end{table}

\textbf{Human Evaluation:} We carried out 12 blind evaluations of 30 information-seeking queries covering mental health (7), politics and policy (6), geography (5), general health (3), legal news (2), and others (4). Each evaluator rated ISQs from the ground-truth dataset (S1), {\fontfamily{cmss}\selectfont ISEEQ-ERL} (S2), and T5-FT CANARD (S3) using Likert score where 1 is the lowest and 5 is the highest. Such evaluations takes huge amount of effort; 4 days were invested for high quality evaluations. A total of 570 ISQs (On average 7 by S1, 7 by S2, and 4 by S3) were evaluated on two guidelines, described in Table \ref{tab:human_eval}. 
We measured their statistical significance by first performing one-way ANOVA and then using Least Significant Difference (LSD) post-hoc analysis \cite{gunaratna2017relatedness}. Across the 30 queries on both guidelines, both S1 and S2 are better (statistically significant) than S3 whereas, even though S2 mean is better than S1, there is no statistical significance between the two systems (we may say they are comparable).

\textbf{Implementation and Training Details:} We implemented our method using Pytorch Lightning on top of the Hugging Face transformer library ~\cite{wolf2019huggingface}. Hyperparameter tuning in {\fontfamily{cmss}\selectfont ISEEQ} is performed using python library ``ray'', setting $\alpha$ = 0.1971 in equation \ref{eq:reward}, $\gamma$ = 0.12 in equation \ref{eq:tune}, and learning rate = 1.17e-5. We train {\fontfamily{cmss}\selectfont ISEEQ} with cross-validation intervals in each epoch, with epochs ranging 100-120 depending on the dataset size. Four NVIDIA Tesla V100 GPUs (16GB) were used for two weeks of training.

\section{Conclusion}
In this research, we introduced, formalized and developed a generic pipeline {\fontfamily{cmss}\selectfont ISEEQ} for generating logically coherent and semantically related ISQs for CIS. {\fontfamily{cmss}\selectfont ISEEQ} outperformed competitive TLM-based baselines in CIS using commonsense knowledge, entailment constraints, and self-guiding through reinforcement learning, trained within a supervised generative-adversarial setting. We established the competency of our method through quantitative experiments and qualitative evaluation on complex discourse datasets. {\fontfamily{cmss}\selectfont ISEEQ} opens up future research directions in CIS by facilitating the automatic creation of large-scale datasets to develop and train improved CIS systems. Crowd-workers can focus on evaluating and augmenting such datasets rather than creating them anew, thus improving dataset standards. The performance of ISEEQ in an online setting was considered beyond the scope of this resource. Broadly construed, through reinforcement learning with the reward on conceptual flow and logical agreement, ISEEQ can be trained to generate questions that are safety constrained and follow a specialized knowledge processing~\cite{sheth2021knowledge}.

\newpage
\bibliography{Main.bib}

\begin{thebibliography}{}

\bibitem[\protect\citeauthoryear{Aliannejadi \bgroup et al\mbox.\egroup
  }{2019}]{aliannejadi2019asking}
Aliannejadi, M.; Zamani, H.; Crestani, F.; and Croft, W.~B.
\newblock 2019.
\newblock Asking clarifying questions in open-domain information-seeking
  conversations.
\newblock In {\em ACM SIGIR}.

\bibitem[\protect\citeauthoryear{Cho \bgroup et al\mbox.\egroup
  }{2021}]{cho2021contrastive}
Cho, W.~S.; Zhang, Y.; Rao, S.; Celikyilmaz, A.; Xiong, C.; Gao, J.; Wang, M.;
  and Dolan, W.~B.
\newblock 2021.
\newblock Contrastive multi-document question generation.
\newblock In {\em Proceedings of the 16th Conference of the European Chapter of
  the Association for Computational Linguistics: Main Volume},  12--30.

\bibitem[\protect\citeauthoryear{Choi \bgroup et al\mbox.\egroup
  }{2018}]{choi2018quac}
Choi, E.; He, H.; Iyyer, M.; Yatskar, M.; Yih, W.-t.; Choi, Y.; Liang, P.; and
  Zettlemoyer, L.
\newblock 2018.
\newblock Quac: Question answering in context.
\newblock In {\em Proceedings of the 2018 Conference on Empirical Methods in
  Natural Language Processing},  2174--2184.

\bibitem[\protect\citeauthoryear{Clark and Manning}{2016}]{clark2016improving}
Clark, K., and Manning, C.~D.
\newblock 2016.
\newblock Improving coreference resolution by learning entity-level distributed
  representations.
\newblock In {\em Proc. ACL}.

\bibitem[\protect\citeauthoryear{Clark \bgroup et al\mbox.\egroup
  }{2019}]{clark2019electra}
Clark, K.; Luong, M.-T.; Le, Q.~V.; and Manning, C.~D.
\newblock 2019.
\newblock Electra: Pre-training text encoders as discriminators rather than
  generators.
\newblock In {\em International Conference on Learning Representations}.

\bibitem[\protect\citeauthoryear{Cohen, Yang, and
  Croft}{2018}]{cohen2018wikipassageqa}
Cohen, D.; Yang, L.; and Croft, W.~B.
\newblock 2018.
\newblock Wikipassageqa: A benchmark collection for research on non-factoid
  answer passage retrieval.
\newblock In {\em The 41st International ACM SIGIR Conference on Research \&
  Development in Information Retrieval},  1165--1168.

\bibitem[\protect\citeauthoryear{Dalton \bgroup et al\mbox.\egroup
  }{2020}]{dalton2020cast}
Dalton, J.; Xiong, C.; Kumar, V.; and Callan, J.
\newblock 2020.
\newblock Cast-19: A dataset for conversational information seeking.
\newblock In {\em Proc. of SIGIR}.

\bibitem[\protect\citeauthoryear{Gao \bgroup et al\mbox.\egroup
  }{2020}]{gao2020discern}
Gao, Y.; Wu, C.-S.; Li, J.; Joty, S.; Hoi, S.~C.; Xiong, C.; King, I.; and Lyu,
  M.
\newblock 2020.
\newblock Discern: Discourse-aware entailment reasoning network for
  conversational machine reading.
\newblock In {\em Proceedings of the 2020 Conference on Empirical Methods in
  Natural Language Processing (EMNLP)},  2439--2449.

\bibitem[\protect\citeauthoryear{Gopalakrishnan \bgroup et al\mbox.\egroup
  }{2019}]{Gopalakrishnan2019}
Gopalakrishnan, K.; Hedayatnia, B.; Chen, Q.; Gottardi, A.; Kwatra, S.;
  Venkatesh, A.; Gabriel, R.; and Hakkani-Tür, D.
\newblock 2019.
\newblock {Topical-Chat: Towards Knowledge-Grounded Open-Domain Conversations}.
\newblock In {\em Proc. Interspeech 2019}.

\bibitem[\protect\citeauthoryear{Gunaratna \bgroup et al\mbox.\egroup
  }{2017}]{gunaratna2017relatedness}
Gunaratna, K.; Yazdavar, A.~H.; Thirunarayan, K.; Sheth, A.; and Cheng, G.
\newblock 2017.
\newblock Relatedness-based multi-entity summarization.
\newblock In {\em Proceedings of the 26th International Joint Conference on
  Artificial Intelligence},  1060--1066.

\bibitem[\protect\citeauthoryear{Guu \bgroup et al\mbox.\egroup
  }{2020}]{guu2020realm}
Guu, K.; Lee, K.; Tung, Z.; Pasupat, P.; and Chang, M.-W.
\newblock 2020.
\newblock Realm: Retrieval-augmented language model pre-training.
\newblock {\em arXiv preprint arXiv:2002.08909}.

\bibitem[\protect\citeauthoryear{Holtzman \bgroup et al\mbox.\egroup
  }{2019}]{holtzman2019curious}
Holtzman, A.; Buys, J.; Du, L.; Forbes, M.; and Choi, Y.
\newblock 2019.
\newblock The curious case of neural text degeneration.
\newblock In {\em International Conference on Learning Representations}.

\bibitem[\protect\citeauthoryear{Karpukhin \bgroup et al\mbox.\egroup
  }{2020}]{karpukhin2020dense}
Karpukhin, V.; Oguz, B.; Min, S.; Lewis, P.; Wu, L.; Edunov, S.; Chen, D.; and
  Yih, W.-t.
\newblock 2020.
\newblock Dense passage retrieval for open-domain question answering.
\newblock In {\em Proceedings of the 2020 Conference on Empirical Methods in
  Natural Language Processing (EMNLP)},  6769--6781.

\bibitem[\protect\citeauthoryear{Kitaev and
  Klein}{2018}]{kitaev2018constituency}
Kitaev, N., and Klein, D.
\newblock 2018.
\newblock Constituency parsing with a self-attentive encoder.
\newblock In {\em Proceedings of the 56th Annual Meeting of the Association for
  Computational Linguistics (Volume 1: Long Papers)},  2676--2686.

\bibitem[\protect\citeauthoryear{Kumar and Callan}{2020}]{kumar2020making}
Kumar, V., and Callan, J.
\newblock 2020.
\newblock Making information seeking easier: An improved pipeline for
  conversational search.
\newblock In {\em Proceedings of the 2020 Conference on Empirical Methods in
  Natural Language Processing: Findings},  3971--3980.

\bibitem[\protect\citeauthoryear{Kusner \bgroup et al\mbox.\egroup
  }{2015}]{kusner2015word}
Kusner, M.; Sun, Y.; Kolkin, N.; and Weinberger, K.
\newblock 2015.
\newblock From word embeddings to document distances.
\newblock In {\em International conference on machine learning},  957--966.
\newblock PMLR.

\bibitem[\protect\citeauthoryear{Lewis \bgroup et al\mbox.\egroup
  }{2020}]{lewis2020retrieval}
Lewis, P.; Perez, E.; Piktus, A.; Petroni, F.; Karpukhin, V.; Goyal, N.;
  K{\"u}ttler, H.; Lewis, M.; Yih, W.-t.; Rockt{\"a}schel, T.; et~al.
\newblock 2020.
\newblock Retrieval-augmented generation for knowledge-intensive nlp tasks.
\newblock {\em arXiv preprint arXiv:2005.11401}.

\bibitem[\protect\citeauthoryear{Li \bgroup et al\mbox.\egroup
  }{2020}]{li2020enriching}
Li, S.; Huang, Z.; Cheng, G.; Kharlamov, E.; and Gunaratna, K.
\newblock 2020.
\newblock Enriching documents with compact, representative, relevant knowledge
  graphs.
\newblock In {\em IJCAI},  1748--1754.

\bibitem[\protect\citeauthoryear{Li \bgroup et al\mbox.\egroup
  }{2021}]{li2021conversations}
Li, Z.; Zhang, J.; Fei, Z.; Feng, Y.; and Zhou, J.
\newblock 2021.
\newblock Conversations are not flat: Modeling the dynamic information flow
  across dialogue utterances.
\newblock {\em arXiv preprint arXiv:2106.02227}.

\bibitem[\protect\citeauthoryear{Lin \bgroup et al\mbox.\egroup
  }{2020}]{lin2020conversational}
Lin, S.-C.; Yang, J.-H.; Nogueira, R.; Tsai, M.-F.; Wang, C.-J.; and Lin, J.
\newblock 2020.
\newblock Conversational question reformulation via sequence-to-sequence
  architectures and pretrained language models.
\newblock {\em arXiv preprint arXiv:2004.01909}.

\bibitem[\protect\citeauthoryear{Michael \bgroup et al\mbox.\egroup
  }{2018}]{michael2018crowdsourcing}
Michael, J.; Stanovsky, G.; He, L.; Dagan, I.; and Zettlemoyer, L.
\newblock 2018.
\newblock Crowdsourcing question-answer meaning representations.
\newblock In {\em Proc. NAACL}.

\bibitem[\protect\citeauthoryear{Pothirattanachaikul \bgroup et al\mbox.\egroup
  }{2020}]{pothirattanachaikul2020analyzing}
Pothirattanachaikul, S.; Yamamoto, T.; Yamamoto, Y.; and Yoshikawa, M.
\newblock 2020.
\newblock Analyzing the effects of" people also ask" on search behaviors and
  beliefs.
\newblock In {\em Proceedings of the 31st ACM Conference on Hypertext and
  Social Media},  101--110.

\bibitem[\protect\citeauthoryear{Pyatkin \bgroup et al\mbox.\egroup
  }{2020}]{pyatkin2020qadiscourse}
Pyatkin, V.; Klein, A.; Tsarfaty, R.; and Dagan, I.
\newblock 2020.
\newblock Qadiscourse-discourse relations as qa pairs: Representation,
  crowdsourcing and baselines.
\newblock In {\em Proc. EMNLP}.

\bibitem[\protect\citeauthoryear{Qi \bgroup et al\mbox.\egroup
  }{2020}]{qi2020prophetnet}
Qi, W.; Yan, Y.; Gong, Y.; Liu, D.; Duan, N.; Chen, J.; Zhang, R.; and Zhou, M.
\newblock 2020.
\newblock Prophetnet: Predicting future n-gram for sequence-to-sequence
  pre-training.
\newblock In {\em Proceedings of the 2020 Conference on Empirical Methods in
  Natural Language Processing: Findings},  2401--2410.

\bibitem[\protect\citeauthoryear{Radlinski and
  Craswell}{2017}]{radlinski2017theoretical}
Radlinski, F., and Craswell, N.
\newblock 2017.
\newblock A theoretical framework for conversational search.
\newblock In {\em Proc. of CIKM}.

\bibitem[\protect\citeauthoryear{Raffel \bgroup et al\mbox.\egroup
  }{2019}]{raffel2019exploring}
Raffel, C.; Shazeer, N.; Roberts, A.; Lee, K.; Narang, S.; Matena, M.; Zhou,
  Y.; Li, W.; and Liu, P.~J.
\newblock 2019.
\newblock Exploring the limits of transfer learning with a unified text-to-text
  transformer.
\newblock {\em arXiv preprint arXiv:1910.10683}.

\bibitem[\protect\citeauthoryear{Rao and Daum{\'e}~III}{2018}]{rao2018learning}
Rao, S., and Daum{\'e}~III, H.
\newblock 2018.
\newblock Learning to ask good questions: Ranking clarification questions using
  neural expected value of perfect information.
\newblock In {\em Proceedings of the 56th Annual Meeting of the Association for
  Computational Linguistics (Volume 1: Long Papers)},  2737--2746.

\bibitem[\protect\citeauthoryear{Reimers and
  Gurevych}{2019}]{reimers2019sentence}
Reimers, N., and Gurevych, I.
\newblock 2019.
\newblock Sentence-bert: Sentence embeddings using siamese bert-networks.
\newblock In {\em Proceedings of the 2019 Conference on Empirical Methods in
  Natural Language Processing and the 9th International Joint Conference on
  Natural Language Processing (EMNLP-IJCNLP)},  3982--3992.

\bibitem[\protect\citeauthoryear{Rodriguez \bgroup et al\mbox.\egroup
  }{2020}]{rodriguez2020information}
Rodriguez, P.; Crook, P.~A.; Moon, S.; and Wang, Z.
\newblock 2020.
\newblock Information seeking in the spirit of learning: A dataset for
  conversational curiosity.
\newblock In {\em Proc. of EMNLP}.

\bibitem[\protect\citeauthoryear{Saeidi \bgroup et al\mbox.\egroup
  }{2018}]{saeidi2018interpretation}
Saeidi, M.; Bartolo, M.; Lewis, P.; Singh, S.; Rockt{\"a}schel, T.; Sheldon,
  M.; Bouchard, G.; and Riedel, S.
\newblock 2018.
\newblock Interpretation of natural language rules in conversational machine
  reading.
\newblock In {\em Proceedings of the 2018 Conference on Empirical Methods in
  Natural Language Processing},  2087--2097.

\bibitem[\protect\citeauthoryear{Sellam, Das, and
  Parikh}{2020}]{sellam2020bleurt}
Sellam, T.; Das, D.; and Parikh, A.
\newblock 2020.
\newblock Bleurt: Learning robust metrics for text generation.
\newblock In {\em Proc. ACL}.

\bibitem[\protect\citeauthoryear{Sheth \bgroup et al\mbox.\egroup
  }{2021}]{sheth2021knowledge}
Sheth, A.; Gaur, M.; Roy, K.; and Faldu, K.
\newblock 2021.
\newblock Knowledge-intensive language understanding for explainable ai.
\newblock {\em IEEE Internet Computing} 25(5):19--24.

\bibitem[\protect\citeauthoryear{Speer, Chin, and
  Havasi}{2017}]{speer2017conceptnet}
Speer, R.; Chin, J.; and Havasi, C.
\newblock 2017.
\newblock Conceptnet 5.5: An open multilingual graph of general knowledge.
\newblock In {\em Thirty-first AAAI conference on artificial intelligence}.

\bibitem[\protect\citeauthoryear{Tarunesh, Aditya, and
  Choudhury}{2021}]{tarunesh2021trusting}
Tarunesh, I.; Aditya, S.; and Choudhury, M.
\newblock 2021.
\newblock Trusting roberta over bert: Insights from checklisting the natural
  language inference task.
\newblock {\em arXiv preprint arXiv:2107.07229}.

\bibitem[\protect\citeauthoryear{Vakulenko, Kanoulas, and de
  Rijke}{2021}]{vakulenko2021large}
Vakulenko, S.; Kanoulas, E.; and de~Rijke, M.
\newblock 2021.
\newblock A large-scale analysis of mixed initiative in information-seeking
  dialogues for conversational search.
\newblock {\em arXiv preprint arXiv:2104.07096}.

\bibitem[\protect\citeauthoryear{Wolf \bgroup et al\mbox.\egroup
  }{2019}]{wolf2019huggingface}
Wolf, T.; Debut, L.; Sanh, V.; Chaumond, J.; Delangue, C.; Moi, A.; Cistac, P.;
  Rault, T.; Louf, R.; Funtowicz, M.; et~al.
\newblock 2019.
\newblock Huggingface's transformers: State-of-the-art natural language
  processing.
\newblock {\em arXiv preprint arXiv:1910.03771}.

\bibitem[\protect\citeauthoryear{Wood \bgroup et al\mbox.\egroup
  }{2018}]{wood2018detecting}
Wood, A.; Rodeghero, P.; Armaly, A.; and McMillan, C.
\newblock 2018.
\newblock Detecting speech act types in developer question/answer conversations
  during bug repair.
\newblock In {\em Proceedings of the 2018 26th ACM Joint Meeting on European
  Software Engineering Conference and Symposium on the Foundations of Software
  Engineering},  491--502.

\bibitem[\protect\citeauthoryear{Wood, Eberhart, and
  McMillan}{2020}]{wood2020dialogue}
Wood, A.; Eberhart, Z.; and McMillan, C.
\newblock 2020.
\newblock Dialogue act classification for virtual agents for software engineers
  during debugging.
\newblock In {\em Proceedings of the IEEE/ACM 42nd International Conference on
  Software Engineering Workshops},  462--469.

\bibitem[\protect\citeauthoryear{Yang \bgroup et al\mbox.\egroup
  }{2018}]{yang2018hotpotqa}
Yang, Z.; Qi, P.; Zhang, S.; Bengio, Y.; Cohen, W.~W.; Salakhutdinov, R.; and
  Manning, C.~D.
\newblock 2018.
\newblock Hotpotqa: A dataset for diverse, explainable multi-hop question
  answering.
\newblock In {\em EMNLP}.

\bibitem[\protect\citeauthoryear{Zamani and Craswell}{2020}]{zamani2020macaw}
Zamani, H., and Craswell, N.
\newblock 2020.
\newblock Macaw: An extensible conversational information seeking platform.
\newblock In {\em Proc. of ACM SIGIR}.

\bibitem[\protect\citeauthoryear{Zamani \bgroup et al\mbox.\egroup
  }{2020}]{zamani2020generating}
Zamani, H.; Dumais, S.; Craswell, N.; Bennett, P.; and Lueck, G.
\newblock 2020.
\newblock Generating clarifying questions for information retrieval.
\newblock In {\em Proceedings of The Web Conference 2020},  418--428.

\bibitem[\protect\citeauthoryear{Zhang and Bansal}{2019}]{zhang2019addressing}
Zhang, S., and Bansal, M.
\newblock 2019.
\newblock Addressing semantic drift in question generation for semi-supervised
  question answering.
\newblock In {\em Proceedings of the 2019 Conference on Empirical Methods in
  Natural Language Processing and the 9th International Joint Conference on
  Natural Language Processing (EMNLP-IJCNLP)},  2495--2509.

\bibitem[\protect\citeauthoryear{Zhang \bgroup et al\mbox.\egroup
  }{2019}]{zhang2019bertscore}
Zhang, T.; Kishore, V.; Wu, F.; Weinberger, K.~Q.; and Artzi, Y.
\newblock 2019.
\newblock Bertscore: Evaluating text generation with bert.
\newblock In {\em ICLR}.

\end{thebibliography}
\bibliographystyle{aaai}

\end{document}